\documentclass[12pt]{iopart}

\usepackage{amssymb}
\usepackage{bm}
\usepackage{graphicx}

\begin{document}

\title[]{Determination of the UV cut-off from the
observed value of the Universe acceleration.}

\author{S.L. Cherkas\dag   \ and V.L. Kalashnikov\ddag
}

\address{\dag\
Institute for Nuclear Problems, Bobruiskaya 11, Minsk 220050,
Belarus}

\address{\ddag\ Technische Universit\"{a}t Wien, Gusshausstrasse 27/387, Vienna A-1040, Austria}

\begin{abstract}
It is shown that use of the equation of motion of the Universe
scale factor allows calculation of the contribution of the vacuum
fluctuations to the acceleration of the Universe expansion.
Renormalization of the equation is needed only in the case of
massive particles. Under a known number of the different kinds of
fundamental fields, this provides the determination of momentum of
the ultraviolet cut-off from the observed value of acceleration.
\end{abstract}


\maketitle
\section{Introduction}

Discovery of the Universe accelerated expansion \cite{ex,ex1} stirs
up theoretical analysis aimed at explanation of the observational
data. On the one hand, most of approaches are based on exploration
of the new kinds of substance (quintessence, Chaplygin's gases,
phantom field, ghost condensate, and so on) or of the new concepts
of space-time (extra-dimensions, holographic principle, string
multiple vacuum) \cite{rev}. On the other hand, it was proposed that
such an explanation can be based on an assumption that Universe or
its constituents are inhomogeneous \cite{rio,but,syk}. The extreme
point of view \cite{rio} is to consider Universe as inhomogeneous at
the super-Hubble scales and to assume that we live at the edge of
``Hubble-bubble'' and thereby can observe either accelerated or
decelerated expansion of Universe, which is not described by the
Friedmann equations. Also, one can assume that the accelerated
expansion is produced by an internal motion of matter on a smaller
scale (like a galaxy formation one \cite{syk}) if the corresponding
averaging procedure is defined for the Universe scale factor
\cite{but}.

Nevertheless, it is still attractive to consider the QFT vacuum in
a non-stationary space-time (i.e. in an expanding Universe) as the
source of the Unverse acceleration. This can be achieved by
exploring of two scales of Universe: $L$ and $1/M_p$. Here $L$ is
the ``size'' of Universe (of the order of its curvature or its
inverse Hubble constant for a flat space-time), $M_p$ is the
Planck mass associated with an appearance of new physics. Whereas
the direct (``naive'') consideration results in the huge value of
vacuum density $\rho_{vac}\sim M_p^4$, the dimensional
regularization of divergent expressions in a manner of Minkowski
QFT\footnote{The works by L. Parker and collaborators
\cite{par,komp} based on the nonperturbative approach stand
apart.} gives the tiny value $\rho_{vac}\sim L^{-4}$
\cite{birrel,one}. In contrast to these approaches, there is a
number of works \cite{coh,pad,pad0,bomb,sred,brust,ver,yar,odn}
obtaining $\rho_{vac}\sim M_p^2L^{-2}$ that is consistent with the
observed value of the Universe acceleration. Ref. \cite{coh} is
simply an observation that the Schwarzschild radius
$\rho_{vac}L^3/M_p^2$ of Universe with the ``mass''
$\rho_{vac}L^3$ has not to exceed the Universe ``size'' $L$. This
excludes the ``naive'' value $\rho_{vac}\sim M_p^4$ and results in
$\rho_{vac}\sim M_p^2L^{-2}$ as an upper limit. Other works are
based on the Zeldovich's idea \cite{zel} assuming that the source
of the Universe acceleration is not in itself the huge QFT vacuum
energy but some another quantity, namely, the square root of
dispersion of the vacuum energy fluctuations.

Although the direct UV cut-off on the Planck level in the
Friedmann equation does not provide the observed value of the
acceleration parameter (and is inconsistent with \cite{coh}), one
can use it in the equation of motion of the Universe scale factor
\cite{prev,prev1}. It is found, that the effective cosmological
constant results from the vacuum fluctuations of the massless
scalar field and its value is concistent with $\rho_{vac}\sim
M_p^2L^{-2}$. The point is that, in contrast to the Friedmann
equation which is held only up to some constant\footnote{ Such a
possibility in the quantum case was pointed out in Ref.
\cite{hawking}. Also, arguments against a prevalence of the
Friedmann equation  were adduced in \cite{prep2}. In this work the
quantization scheme for the equation of motion is proposed
resulting in the quasi-Heisenberg operators which act in a space
of solutions of the Wheeler-DeWitt equation. The hamiltonian
constraint (i.e. the Friedmann equation) appears from two points
of view there. The first one is properly the Wheeler-DeWitt
equation and the second one is the relation for the
quasi-Heisenberg operators. In the second sense, the constraint
(Friedmann equation) is broken during the Universe evolution, when
the potential energy of the scalar field begins to play a role.}
in such an approach, the equation of motion contains difference of
the potential and kinetic energies of the field oscillators. In
Minkowski space-time this difference is exactly zero, because the
virial theorem for an oscillator states that the kinetic energy is
equal to the potential one in the virial equilibrium. However,
 in the expanding Universe the difference is proportional to the Hubble
constant squared. This last tiny quantity is multiplied by the
Planck-order momentum cut-off squared that gives true order of the
acceleration parameter.

The aim of this work is to analyze a number of fundamental fields,
which can contribute to the cosmological constant, and
 to investigate a possibility
of exact determination of the UV cut-off from the observed value of
the acceleration parameter.

\section{Contribution of the scalar particles}

In the case of the homogeneous isotropic flat Universe filled with
the different species of matter, Lagrangian has the form
\cite{prev}:
\begin{equation}
\mathcal{ L}=-\frac{M_p^2 {a^\prime}^2}{2}+\mathcal{
L}_{scal}+\mathcal{ L}_{ferm}+\dots,
\end{equation}
where the metric $ds^2=a^2(\eta)(d\eta^2-d\bm r^2)$ is assumed and
$M_p=\sqrt{\frac{3}{4 \pi G}}$ is the Planck mass. Matter
Lagrangians are
\begin{equation}
\mathcal{ L}_{scal}=\frac{1}{2}\int \biggl(a^2
(\partial_\eta\phi)^2- a^2 (\bm \nabla \phi)^2-a^4 m^2\phi^2
\biggr)d^3\bm r, \label{sc}
\end{equation}

\begin{equation}
\mathcal{ L}_{ferm}=\int \biggl(\frac{i\, a^3
}{2}\psi^+\partial_\eta\psi-\frac{i\, a^3}{2}
\partial_\eta \psi^+\psi+i a^3\psi^+\bm \alpha \bm \nabla\psi-a^4\, m \psi^+\beta
\psi\biggr)d^3\bm r, \label{ff}
\end{equation}
where $\psi(\eta,\bm r)$ is the spinor field and $\phi(\eta,\bm
r)$ is the scalar field and three dimensional volume of
integration is assumed to be unity.

Let us turn to the  Fourier-transform
 $\phi(\bm
r)=\sum_{\bm k} \phi_{\bm k} e^{i {\bm k}\bm r}$  that leads to
the scalar field Lagrangian in the form

\begin{equation}
\mathcal{ L}_{scal}=\frac{1}{2}\sum_{\bm k} a^2{\phi^\prime_{\bm
k}\phi^\prime_{-{\bm k}}}- a^2 k^2 \phi_{\bm k}\phi_{-{\bm k}} -
a^4 m^2 \phi_{\bm k}\phi_{-{\bm k}},.
\end{equation}
In the quasi-classical picture, where the Universe scale factor is
considered classically whereas the scalar field is quantum,
equations of motion are:

\begin{eqnarray}
\hat\phi^{\prime\prime}_{\bm k}+(k^2+a^2 m^2)\hat\phi_{\bm
k}+2\frac{
a^\prime}{a}{ \hat \phi^\prime}_{\bm k}=0,\\
M_p^2\,a^{\prime\prime}+\frac{1}{a}\sum_{\bm k} a^2<\hat
\phi^\prime_{\bm k}\hat\phi^\prime_{-{\bm k}}> -a^2
k^2<\hat\phi_{\bm k}\hat\phi_{-{\bm k}}> \nonumber\\-2 a^4 m^2
<\hat \phi_{\bm k}\hat\phi_{-{\bm k}}>=0. \label{sys}
\end{eqnarray}
\noindent Here $<>$ denotes averaging over the vacuum state.

Quantization of the scalar field \cite{birrel}
\begin{equation}
\hat \phi_{\bm k}=\hat {\mbox{a}}^+_{-\bm k}\chi_{k}^*(\eta)+\hat
{\mbox{a}}_{\bm k} \chi_{k}(\eta)
\end{equation}
leads to the operators of creation and annihilation with the
commutation rules
 $[{\hat{\mbox{a}}}_{\bm k},\, {\hat{\mbox{a}}}^+_{\bm k}]=1$. The complex functions
 $\chi_k(\eta)$ are
 $\chi_k=\frac{1}{a\sqrt{2 k}}\,e^{-ik\eta }$ for $a=const$.

 In the general case
 they satisfy the relations \cite{birrel}:
\begin{eqnarray}
\chi^{\prime\prime}_k+k^2 \chi_k+2\frac{ a^\prime}{a}{
\chi^\prime}_k=0,\nonumber\\
a^2(\eta)(\chi_k \,{\chi_k^\prime}^*-\chi_k^*\,\chi_k^\prime)=i.
\label{rel}
\end{eqnarray}

Eqs. (\ref{rel}) admit the formal WKB solution \cite{birrel}

\begin{equation}
 \chi_{k}=\frac{Exp\left(-i \int_0^\eta W_k(\tau) \, d\tau \right)}{\sqrt{2} a(\eta)
\sqrt{W_k(\eta)}},
\end{equation}
where function $W_k(\eta)$ satisfies the equation
\begin{equation}
W_k''-\frac{3 W_k'^2}{2
   W_k}-2 \left(k^2+m^2 a^2-\frac{a^{\prime\prime}}{a}\right) W_k+2 W_k^3=0.
\label{ad}
\end{equation}
Adiabatic approximation consists in setting $W_k(\eta)\approx
\sqrt{k^2+m^2 a(\eta )^2-a''(\eta)/a(\eta)}$.

At first, we shall consider the massless case $m=0$ in which the
equation of motion contains exact difference of the kinetic and
potential energies of the field oscillators. This difference is zero
in the vacuum state of Minkowski space. In the expanding Universe it
takes the form
\begin{eqnarray}
a^2<\hat \phi^\prime_{\bm k}\hat\phi^\prime_{-{\bm k}}>- k^2a^2
<\hat\phi_{\bm k}\hat\phi_{-{\bm k}}>=
a^2({\psi_k^\prime}^{*}\psi_k^\prime-k^2\psi_k^*
\psi_k)\nonumber\\\approx \frac{1}{2
k}\left(-\frac{a^{\prime\prime}}{a}+\frac{{a^\prime}^2}{a^2}\right)
+O({a^\prime}^3)+O(a^\prime
a^{\prime\prime})+O(a^{\prime\prime\prime}), \label{15}
\end{eqnarray}
when the adiabatic approximation for the function $\psi_k$ is used.
Using (\ref{15}) in (\ref{sys}) yields the equation of motion for
the Universe scale factor in the form
\begin{equation}
M_p^2\,a^{\prime\prime}+\frac{1}{2 a
}\left(-\frac{a^{\prime\prime}}{a}+\frac{{a^\prime}^2}{a^2}\right)\sum_{\bm
k}\frac{1}{k} =0.
\end{equation}
Coming from the summation over $\bm k$ (in the general case it has
to include summation over particle species) to the integration, we
have
\begin{equation}
\sum_{\bm k} \frac{1}{k}=\frac{4 \pi}{(2\pi)^3}\int_0^{k_{max}} k
\,dk = \frac{k_{max}^2}{4 \pi^2 }=\kappa_{max}^2\frac{a_0^2}{4
\pi^2 },
\end{equation}
where $\kappa_{max}=\frac{k_{max}}{a_0}$ is the cutting parameter
of the physical momentums and $a_0=a(0)$ is the present day
($\eta=0$) scale factor. Finally, we come to the equation

\begin{equation}
M_p^2\,a^{\prime\prime}=\rho \,a^3 -\frac{1}{2 a
}\left(-\frac{a^{\prime\prime}}{a}+\frac{{a^\prime}^2}{a^2}\right)N_{sc}\kappa_{max}^2\frac{a_0^2}{4
\pi^2 }, \label{14}
\end{equation}
where the density of classical matter (dust) $\rho=\frac{\rho_0
a_0^3}{a^3}=\frac{\Omega_m M_p^2\,\mathcal H^2 a_0}{2\, a^3}$ and
the number $N_{sc}$ of species of scalar particles are introduced.
Here $\Omega_m=\rho_0/\rho_{crit}$ should be read as denoting  the
matter relative density  and $\mathcal
H=\frac{a^\prime}{a}|_{\eta=0}$.

Consideration of Eq. (\ref{14}) at $\eta=0$ allows determining the
UV cut-off $\kappa_{max}$ of physical momentum from the observed
present day value of the Universe acceleration
\begin{equation}
\kappa_{max}=\frac{2\pi M_p}{\sqrt{N_{sc}}}
\sqrt{\frac{2a^{\prime\prime}a/{a^\prime}^2-\Omega_m}{a^{\prime\prime
}a/{a^\prime}^2-1}}\,\,\Biggr|_{\eta=0}. \label{cut}
\end{equation}

Now let us return to the massive particles. In this case there is no
an exact difference of the kinetic and potential energies in the
equation of motion because the multiplier $2$ appears in Eq.
(\ref{sys}). Subtraction of the main terms corresponding to the
Minkowski space-time can be used in this case as a primitive
renormalization. Let us remind that for $a=const$
\begin{eqnarray*}
{a^4}<\phi^\prime_{k}\phi^\prime_{-k}>= <\pi_k\pi_{-k}>=\frac{1}{2}a^2\sqrt{k^2+m^2a^2},\\
<\phi_k\phi_{-k}>=\frac{1}{2\,a^2\sqrt{k^2+m^2a^2}}.
\end{eqnarray*}
The renormalization consists in rewriting of the equation of
motion in the form
\begin{equation}
a^{\prime\prime}M_p^2-\rho a^3 =-\sum_k
\frac{\theta_k}{a^3}-a(k^2+2a^2m^2)\chi_k,
\end{equation}
where
\begin{eqnarray}
 \theta_k=<\pi_k\pi_{-k}>-\frac{1}{2}a^2\sqrt{k^2+m^2a^2},\nonumber\\
\chi_k=<\phi_k\phi_{-k}>-\frac{1}{2\,a^2\sqrt{k^2+m^2a^2}}.
\label{renorm}
\end{eqnarray}

The simplest way of further consideraton  is to assume some
particular dependence $a(\eta)$ and to calculate the quantity
$\sum_k \frac{\theta_k}{a^3}-a(k^2+2a^2m^2)\chi_k$. As it was done
in \cite{prev}, one can choose the dependence
\begin{equation}
a(\eta)=a_0(1 - \mathcal{H}\eta/2  )^{-2}, \label{aeta}
\end{equation}
which can be roughly associated with our present Universe because
it gives the acceleration parameter
$a^{\prime\prime}a/{a^\prime}^2=3/2$ and
$a^\prime/a\bigr|_{\eta=0}=\mathcal H$ (here $\mathcal H=H a_0 $
and $H$ is the Hubble constant).

Calculation of the scalar field contribution to the Universe
acceleration up to second order in $\mathcal H$ gives

\begin{eqnarray}
a^{\prime\prime}M_p^2-\rho a^3= -\sum_k
\frac{\theta_k}{a^3}-a(k^2+2a^2m^2)\chi_k\nonumber\\\approx\sum_k
\frac{\mathcal{H}^2 k^4}{4 a_0 \left(k^2+m^2 a_0^2
\right)^{5/2}}+\frac{3 a_0 \mathcal{H}^2 m^2 k^2}{8
\left(k^2+m^2a_0^2\right)^{5/2}}+O(\mathcal
H^4)\nonumber\\\approx\sum_{\bm k}\frac{\mathcal H^2}{4 a_0
k}-\frac{a_0
\mathcal H^2 m^2}{4 k^3}+O(m^4)+O(\mathcal H^4)\nonumber\\
\approx\frac{a_0\mathcal H^2 \kappa_{max}^2}{16
\pi^2}\left(1-\frac{2 m^2
}{\kappa_{max}^2}\ln\frac{\kappa_{max}}{\kappa_{min}}\right),
\end{eqnarray}
where replacement of summation by integration over the momentum
$k$ is proceeded and $\kappa_{min}\sim m $. One can see, that the
mass terms are suppressed by the multiplier $m/\kappa_{max}\sim
m/M_p$ and are negligible, if one assumes that the mass of a
particle is mach smaller than the Planck mass. Thus, one can
conclude that the mass of the scalar particles has no impact.

\section{Fermions}

To consider the fermions in an expanding Universe, we keep to the
earlier work \cite{el} and the works \cite{kof,pel}, where
preheating of Universe was considered (see also \cite{maa,zg}).

After decomposing in the complete set of modes
 $\psi(\bm
r)=\sum_{\bm k} \psi_{\bm k} e^{i {\bm k}\bm r}$, the fermionic
Lagrangian (\ref{ff}) takes the form
\begin{equation}
\fl \mathcal{ L}_{ferm}=\sum_{\bm k}\frac{i\, a^3 }{2}\psi^+_{\bm
k}\partial_\eta\psi_{\bm k}-\frac{i\, a^3}{2}
\partial_\eta \psi^+_{\bm k}\psi_{\bm k}-a^3\psi^+_{\bm k}
(\bm \alpha \bm k)\,\psi_{\bm k}-a^4\, m \psi^+_{\bm k}\beta
\psi_{\bm k}.
\end{equation}

The equations of motion are
\begin{eqnarray}
i{\hat\psi}^\prime_k- (\bm \alpha  \bm k){\hat\psi}_{\bm k} +i
\frac{3a^\prime}{2a}{\hat\psi}_{\bm k}-m\,a \beta{\hat\psi}_{\bm k}=0,\nonumber\\
m_p^2a^{\prime\prime}+ \sum_{\bm k}\frac{3}{2}a^2
<i{\hat\psi}^+_{\bm k}{\hat\psi}^\prime_{\bm
k}-i{\hat\psi}^{\prime +}_{\bm k}{\hat\psi}_{\bm k}-2
{\hat\psi}^+_{\bm k}(\bm \alpha \bm k){\hat\psi}_{\bm
k}>\nonumber\\-4 a^3 m<{\hat\psi}^+_{\bm k}\beta{\hat\psi}_{\bm
k}>+{\bf \dots}=0, \label{rr}
\end{eqnarray}
where quasi-classics is implied and dots denote matter (dust),
scalar field, photon field and other terms.

Fermion field is quantized as
\begin{equation}
\hat \psi_{\bm k}={\hat b}^+_{-\bm k,s}v_{-\bm k,s}+{\hat a}_{\bm
k,s}u_{\bm k,s},
\end{equation}
where the bispinor is\footnote{Representation of the Dirac matrices
is the same as in Refs. \cite{lan4,cher}.}:
\[u_{\bm k,s}(\eta) = \frac{i\chi_k^\prime+m a\chi_k}{a^{3/2}}\left (
\begin {array}{c}
 \varphi_s \\
\frac{\chi_k(\bm \sigma \bm k)}{i\chi_k^\prime+m \chi_k
a}\varphi_s
\end {array}
\right)~,
\]
 $\varphi_+=\left (
\begin {array}{c}
1 \\
0
\end {array}
\right)$ and $\varphi_-=\left (
\begin {array}{c}
0 \\
1
\end {array}
\right)$.

The bispinor $v_{\bm k,s}$  is expressed as  $v_{\bm
k,s}=i\gamma^0\gamma^2(\bar u_{\bm k,s})^{T}$, where the symbol $T$
denotes the transpose vector and $\bar u=u^+\gamma^0$. The functions
$\chi_{k}(\eta)$ satisfy the equations

\begin{eqnarray}
\chi^{\prime\prime}_{k }+ \left(k^2+m^2 a^2-i m a'\right)\chi_{
k}=0, \label{1}
\\
 k^2\chi_{ k}  \chi^*_{ k} +\left(a m \chi^*_{k}-i {\chi
^\prime_{k}}^*\right) \left(a m \chi_{k} +i \chi^\prime_{ k}
\right)=1. \label{rel1}
\end{eqnarray}
Eq. (\ref{1}) reflects the time-dependence dictated by the
bispinor equation of motion. Eq. (\ref{rel1}) is required to
satisfy the anti-commutator relation $\{{\psi^{+j}_{\bm
k,s}},\psi^l_{\bm q,\sigma}\}_{+}=\delta_{j l}\delta_{\bm k \bm
q}\delta_{s \sigma}$, so that the operators $a_{\bm  k, s}$
satisfy $\{a_{\bm k,s}a^+_{\bm q,\sigma}\}_+=\delta_{\bm k \bm
q}\delta_{s \sigma}$ and the similar relation is valid for $b_{\bm
k,s}$. Also, one has to note that the left hand side of Eq.
(\ref{rel1}) is the integral of motion of Eq. (\ref{1}).

Eq. (\ref{1}) admits the WKB solution:
\begin{equation}
\chi_{k}(\eta)=\frac{\exp\left({-i \int_0^{\eta } W_k(\tau) \,
d\tau-\frac{1}{2} \int_0^{\eta } \frac{m a'(\tau)}{W_k(\tau)} \,
d\tau}\right)}{\sqrt{2} \left(m a_0+\sqrt{k^2+m^2
   a_0^2}\right)^{1/2} \sqrt{W_k(\eta )}},
\end{equation}
where the function $W_k(\eta)$ satisfies the equation:
\begin{equation}
W_k''-\frac{1}{2W_k}\left({m^2 a'^2}+{3 W_k'^2}+{4 m a'
W_k'}\right)+2 W_k^3-2 (k^2 + m^2 a^2) W_k+m a''=0. \label{wkf}
\end{equation}

It should be noted, that Eq. (\ref{wkf}) gives $W_k=k$ at $m=0$
and this can be seen directly from Eqs. (\ref{1}), (\ref{rel1})
giving $\chi_k(\eta)=\frac{1}{\sqrt{2}\,k}e^{-ik\eta }$. In this
case, the elementary calculations lead to
\begin{eqnarray}
\frac{3}{2}a^2 <i{\hat\psi}^+_{\bm k}{\hat\psi}^\prime_{\bm
k}-i{\hat\psi}^{\prime +}_{\bm k}{\hat\psi}_{\bm k}-2
{\hat\psi}^+_{\bm k}(\bm \alpha \bm k){\hat\psi}_{\bm
k}>\nonumber\\ =\frac{3}{a}({ i \chi_k^{*} \chi_k ' k^2}-{ i
\chi_k \chi_k^{*\prime}k^2}-{ i \chi_k^{*\prime} \chi_k ''}+{ i
\chi_k ' \chi_k^{*\prime\prime}})=0.
\end{eqnarray}
 Thus, according to (\ref{rr}) the massless fermions do not contribute to the Universe
acceleration.

To analyze the massive fermions, let us assume the particular
dependence $a(\eta)$ given by (\ref{aeta}) and calculate the
quantity
\begin{eqnarray}
\sum_{\bm k}\frac{3}{2}a^2 <i{\hat\psi}^+_{\bm
k}{\hat\psi}^\prime_{\bm k}-i{\hat\psi}^{\prime +}_{\bm
k}{\hat\psi}_{\bm k}-2 {\hat\psi}^+_{\bm k}(\bm \alpha \bm
k){\hat\psi}_{\bm k}>-4 a^3 m<{\hat\psi}^+_{\bm
k}\beta{\hat\psi}_{\bm k}>\nonumber\\
=\sum_{\bm k}\frac{2 {a_0} m^2}{k}-\frac{3 {a_0} m^2 {\mathcal
H}^2}{k^3}-\frac{9 m {\mathcal H}^2}{4 k^2}+O(m^3)+O(\mathcal
H^4)\nonumber\\
\approx \frac{a_0^3 \kappa_{max}^2m^2}{2 \pi^2}-\frac{9
a_0\mathcal H^2 \kappa_{max}^2 }{8
\pi^2}\left(\frac{m}{\kappa_{max}}+\frac{4
m^2}{3\kappa_{max}^2}\ln\frac{\kappa_{max}}{\kappa_{min}}\right)~
\end{eqnarray}
contained in Eq. (\ref{rr}). One can see that after subtraction of
the first term (which is equivalent to the ``renormalization''
(\ref{renorm}) in the case of the scalar field), the remaining terms
are suppressed by the multiplier $m/\kappa_{max}$ in comparison with
the acceleration term created by the vacuum of the massless scalar
field.

It seems that the negligible contribution of the massive terms is a
general feature. Indeed, the quantity $M_p^2a^{\prime\prime}/a$ is
of the forth-order in mass. From the dimension arguments, it can be
proportional to $\mathcal H^2\kappa_{max}^2$, $\mathcal
H^2\kappa_{max} m$, $\mathcal H^2 m^2$ and two last terms are
negligible in comparison with the first one at $\kappa_{max}\sim
M_p$. Note, that the term $\mathcal H m \kappa_{max}^2$ is forbidden
by the time-reversal invariance arising from the fluctuation origin
(i.e. the vacuum fluctuations) of the quantity under consideration.

\section{Conformally coupled scalar fields, photons and any conformal fields}

As one can see, that the massless fermions do not contribute to the
Universe acceleration. The deep-laid reason (see for instance
\cite{pr}) is conformality of the fermionic Lagrangian if $m=0$.
Other conformal Lagrangians correspond to both photonic and scalar
fields with the $R/6$ term for the latter \cite{birrel}. For
shortness it is sufficiently to consider only the last Lagrangian,
which takes the form:
\begin{equation}
{\mathcal L}_{conf}=-\frac{1}{2}{M_p^2 a^\prime}^2
+\frac{1}{2}\sum_{\bm k} \{y^\prime_{\bm k} y^\prime_{-{\bm
k}}-m^2a^2\, y_{\bm k} y_{-\bm k}-k^2y_{\bm k} y_{-\bm k}\},
\end{equation}
where $y_{\bm k}=a(\eta)\phi_{\bm k}$. One can see, that the field
oscillators are uncoupled from the Universe scale factor in the case
of $m=0$ and, thereby, do not contribute to the equation of motion
for it. Equivalent conclusion, that the massless conformal scalar
field does not produce a vacuum polarization has been done in
\cite{star2}.

\section{Gravitons}
Let us turn to the pure gravity. Small oscillations of the graviton
oscillators near an equilibrium point can be considered on the basis
of the perturbation theory \cite{lifsh}. It is known that the scalar
and vector modes cannot produce free oscillations without the matter
fields \cite{lifsh}. On the other hand, it is possible to expect
that the matter fields are uncoupled from the scalar and vector
perturbations at the Planck frequencies, that is the frequencies
providing main contribution to the vacuum energy within the
framework of our approach.

Thus, the remaining perturbations are the tensor ones existing in
an empty space. Let us represent metric in the form\footnote{We
are going to do calculations up to the second-order in $h$, thus
the quadratic terms on $h$ are kept in the metric.}
\begin{equation}
g_{\mu\nu}=\gamma_{\mu\beta}\left(\delta^\beta_\nu+h^\beta_\nu+\frac{1}{2}h^\beta_\rho
h^\rho_\nu \right), \label{metr}
\end{equation}
where $\gamma _{\mu\nu}=a^2(\eta)\mbox{diag}\{1,-1,-1,-1\}$. All
components of the perturbation $h^\mu_\nu(\eta,\bm r)$ containing at
least one zero index are equal to zero and the spatial
 components
are represented as
\begin{equation}
h^j_n(\eta,\bm r)=\frac{2\sqrt{6}}{M_p}\sum_{\bm
k}\biggl(\bigl[\Phi_{\bm k}\bigr]^j_n\,\phi_{\bm
k}(\eta)+\bigl[\mathcal Y_{\bm k}\bigr]^j_n \,y_{\bm
k}(\eta)\biggr)e^{i \bm k \bm r}. \label{m2}
\end{equation}
The symmetric tensors  $\Phi_{\bm k}$ and  $\mathcal Y_{\bm k}$
describing two possible polarizations  have the following
properties: $Tr[\Phi_{\bm k}]=Tr[\mathcal Y_{\bm k}]=0$,
$~~Tr[\mathcal Y_{\bm k}\,\mathcal Y_{-\bm k}]=Tr[\Phi_{\bm
k}\,\Phi_{-\bm k}]=1$, $~~\Phi_{\bm k}\cdot \bm k=\mathcal Y_{\bm k
}\cdot \bm k=0$. For instance, when the $\bm k$-vector is directed
along $x$-axis, these tensors look as:
\[\mathcal Y_{\bm k} = \frac{1}{\sqrt{2}}\left (
\begin {array}{ccc}
 0 &0&0\\
 0 &1&0\\
 0 &0&-1\\
\end {array}
\right)~,~~~~~~ \Phi_{\bm k} = \frac{1}{\sqrt{2}}\left (
\begin {array}{ccc}
 0 &0&0\\
 0 &0&1\\
 0 &1&0\\
\end {array}
\right).
\]
For an arbitrary direction, they can be obtained from these
expressions through the SO(3)-transformation.

The next step is to substitute  Eqs. (\ref{metr}), (\ref{m2}) in
the gravity action $S=-\frac{M_p^2}{12}\int \mathcal
G\sqrt{-g}\,d^4x$, where $\mathcal G=g^{\alpha \beta }
\left(\Gamma _{\alpha \nu }^{\rho }
   \Gamma _{\beta \rho }^{\nu }-\Gamma _{\alpha \beta
   }^{\nu } \Gamma _{\nu \rho }^{\rho }\right)$ according to
   \cite{lan2}. Finally, in the second-order in $y_{\bm
   k}$, $\phi_{\bm k}$ we come to

\begin{equation}
\fl S=\int \left(-\frac{M_p^2
{a^\prime}^2}{2}+\frac{1}{2}\sum_{\bm k} a^2{\phi^\prime_{\bm
k}\phi^\prime_{-{\bm k}}}- a^2 k^2 \phi_{\bm k}\phi_{-{\bm
k}}+a^2{y^\prime_{\bm k} y^\prime_{-{\bm k}}}- a^2 k^2 y_{\bm
k}y_{-{\bm k}}\right)d\eta.
\end{equation}

Thus the tensor wave contributes to the vacuum energy as two
scalar fields.
\section{Conclusion}

In the case of the massless particles and when the conformal time
is used, it is shown that the equation of motion of the Universe
scale factor contains the difference of potential and kinetic
energies of field oscillators. Calculation of the mean value of
this quantity over the vacuum state leads to the Universe
accelerated expansion. The observed value of the Universe
acceleration allows determining the UV cut-off of the physical
momentum given by Eq. (\ref{cut}), which contains the dust density
and the number of species
 of scalar fields $N_{sc}$.

\begin{figure}[h]
\vspace{0.6 cm} \hspace{2.5 cm}
 \includegraphics[width=3.5 cm]{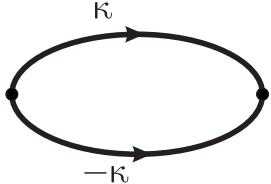}
 \vspace{. cm}
\caption{ Diagram of vacuum polarization. Bold lines denote a
virtual particle propagation in a background metric of expanding
Universe.} \label{diag}
\end{figure}

Let us make some notes about the number of species of scalar
particles. At first sight, there appear to  exists a number of the
scalar particles like pions. However, the momentum $\kappa$ of a
virtual particle appears in the loop diagram and the main
contribution arises from $\kappa$ of the Planck mass order. Under
the circumstances, the composite particles decompose into their
constituents. Thus, $N_{sc}$ should be read as a number of the
fundamental scalar fields.

In a minimal variant of the standard model there is the SU(2)
duplet of complex scalar fields \cite{stand}, i.e. four scalar
degrees of freedom. Adding two degrees of freedom of the tensor
gravitational wave, one comes to $N_{cs}=6$ and the substitutions
of $a^{\prime\prime}a/{a^\prime}^2=3/2$ and $\Omega_m=1/3$ in Eq.
(\ref{cut}) lead to the UV cut-off of the physical momentum:
$\kappa_{max}=\frac{4\sqrt{2} \pi M_p}{ 3}=2\sqrt{\frac{2\pi}{3
G}}$.

Still, the scale of the Grand Unfed Theory $10^{15}-10^{16}$ GeV
seems closer to the Planck scale and, hence there are ten scalar
degrees of freedom  for SU(5) (i.e. total twelve). Then
$\kappa_{max}=2\sqrt{\frac{\pi}{3 G}}$.

Scalar sectors of both the string theory and
      the supersymmetry one providing
a number of the scalar superpartners of fermions demand a deeper
insight.

Let us remind that photons, fermions and conformally coupled scalars
do not contribute to the Universe acceleration. The masses of all
particles gives a negligible contribution after separation of the
term corresponding to the Minkowski space.

\end{document}